\documentclass[aps,prl,reprint,twocolumn,superscriptaddress,preprintnumbers,nofootinbib]{revtex4-1}

\usepackage{amsthm}
\usepackage{amsmath}
\usepackage{graphicx}
\usepackage{subcaption}
\usepackage{slashed}
\usepackage{amssymb}
\usepackage{float}
\usepackage[utf8]{inputenc}
\usepackage[T1]{fontenc}
\usepackage[colorlinks=True, citecolor=blue, urlcolor=blue, linkcolor=blue]{hyperref}

\newcommand*\diff{\mathop{}\!\mathrm{d}}

\newcommand{\nn}{\nonumber}

\newcommand{\be}{\begin{eqnarray}}
\newcommand{\ee}{\end{eqnarray}}

\newcommand{\ml}{\mathcal}
\newcommand{\bs}{\boldsymbol}
\newcommand{\Tr}{\mathrm{Tr}}

\begin{document}

\title{Gauge Invariance of Non-Abelian Field Strength Correlators: the Axial Gauge Puzzle}

\author{Bruno Scheihing-Hitschfeld}
\email{bscheihi@mit.edu}
\affiliation{Center for Theoretical Physics, Massachusetts Institute of Technology, Cambridge, Massachusetts 02139, USA}

\author{Xiaojun Yao}
\email{xjyao@mit.edu}
\email{xjyao@uw.edu}
\affiliation{Center for Theoretical Physics, Massachusetts Institute of Technology, Cambridge, Massachusetts 02139, USA}
\affiliation{InQubator for Quantum Simulation, University of Washington, Seattle, Washington 98195, USA}

\date{\today}
\preprint{MIT-CTP/5425,~IQuS@UW-21-040}
\begin{abstract}
Many transport coefficients of the quark-gluon plasma and nuclear structure functions can be written as gauge invariant correlation functions of non-Abelian field strengths dressed with Wilson lines.
We discuss the applicability of axial gauge $n\cdot A=0$ to calculate them. In particular, we address issues that appear when one attempts to trivialize the Wilson lines in the correlation functions by gauge-fixing. We find it is always impossible to completely remove the gauge fields $n\cdot A$ in Wilson lines that extend to infinity in the $n$-direction by means of gauge transformations. We show how the obstruction appears in an explicit example of a perturbative calculation, and we also explain it more generally from the perspective of the path integral that defines the theory. Our results explain why the two correlators that define the heavy quark and quarkonium transport coefficients, which are seemingly equal in axial gauge, are actually different physical quantities of the quark-gluon plasma and have different values. Furthermore, our findings provide insights into the difference between two inequivalent gluon parton distribution functions.
\end{abstract}

\maketitle

\textbf{Introduction.} Gauge theory plays an essential role in the development of modern physics, highlighted in the formulation of the Standard Model of particle physics
~\cite{Itzykson:1980rh,peskin1995introduction,Weinberg:1996kr,Burgess:2006hbd,Srednicki:2007qs,Schwartz:2014sze}. Besides high energy and particle physics, gauge theory also has wide applications in studies of condensed matter physics~\cite{kleinert1989gauge,fradkin2013field}.
Formally, a gauge theory is specified by a gauge group under which the matter fields and force carriers (gauge fields) transform, and described by a Lagrangian density that is invariant under local gauge transformations.

The gauge symmetry corresponds to a redundancy in the degrees of freedom of the theory, which causes difficulties in quantizing the theory. The most widely employed method to overcome the problem is the Faddeev-Popov (FP) path integral approach~\cite{Faddeev:1967fc}. In the FP quantization, one chooses a gauge condition to remove the redundancy in the gauge field degrees of freedom, obtaining different Lagrangian densities for each gauge choice. Calculations with different gauge choices lead to the same results for physical observables, since they are experimentally measurable and thus gauge invariant quantities. 


A particular gauge choice, called axial gauge, which sets one component of the gauge field to zero $n^\mu A_\mu=0$,\footnote{Here $n^\mu$ is a fixed 4-vector. Our definition of axial gauge is general and includes temporal axial gauge ($n^2>0$), spatial axial gauge ($n^2<0$) and light-cone gauge ($n^2=0$).} has been widely investigated~\cite{Schoenmaker:1981eq,Ball:1981vd,Caracciolo:1982dp,West:1982gg,Landshoff:1985fv,Cheng:1986hv,Leibbrandt:1987qv,James:1990fd,Nyeo:1991if,Joglekar:1999zt}.
However, the use of axial gauge has often led to confusing and seemingly inconsistent results that lack simple physical interpretation due to extra prescriptions required in the calculation. These subtleties become unavoidable in multi-loop calculations for time-ordered quantities~\cite{Leibbrandt:1987qv,Nyeo:1991if}. A famous example where axial gauge causes a subtlety is the transverse momentum dependent parton distribution function (TMD)~\cite{Boer:2011fh,Angeles-Martinez:2015sea,Shanahan:2019zcq,Ebert:2022cku,Ebert:2022fmh}. In light-cone gauge, the gauge links along the light-cone direction in TMDs become trivial and it is essential to include a transverse gauge link at $x^-=\infty$~\cite{Ji:2002aa},\footnote{This corresponds to the infinite light-cone time for a parton moving along the $-z$ direction, such as the struck quark in deep inelastic scattering~\cite{Brodsky:2002cx}.} which has different physical interpretations depending on the boundary conditions of the gauge fields. Different boundary conditions give different prescriptions in the axial gauge gluon propagator but the final result is the same~\cite{Belitsky:2002sm}.
Another example is the transport coefficients of heavy quarks~\cite{Casalderrey-Solana:2006fio,Caron-Huot:2009ncn} and quarkonia~\cite{Brambilla:2016wgg,Brambilla:2017zei,Yao:2020eqy}, which govern their dynamics in the quark-gluon plasma (QGP), a nearly perfect fluid produced in relativistic heavy ion collisions.
A comparison between the perturbative calculations of two correlation functions that define the heavy quark~\cite{Burnier:2010rp} and quarkonium~\cite{Binder:2021otw} transport coefficients, suggests that axial gauge can raise consistency issues even at next-to-leading order (NLO) for gauge invariant correlation functions that involve Wilson lines of infinite extent. In particular, Feynman gauge calculations show these two correlation functions differ in values, but they look identical in temporal axial gauge~\cite{Binder:2021otw}. This is the axial gauge puzzle we want to address in this letter.\footnote{Ref.~\cite{Eller:2019spw} noted axial gauge could be problematic, but did not explicitly address it.}

We will illuminate the origin of the difficulties in applying axial gauge to calculate these quantities (TMDs and QGP transport coefficients), which are defined through correlation functions of the field strength tensors $F_{\mu \nu} \equiv F_{\mu \nu}^a T^a_F$ dressed with Wilson lines. Here $T^a_F$ denote the generators of the SU($N_c$) gauge group in the fundamental representation that satisfy color trace normalization ${\rm Tr}_c(T_F^aT_F^b)=T_F\delta^{ab}$. Proper Wilson lines are necessary for gauge invariance, since in non-Abelian gauge theories, the field strength transforms as $F_{\mu \nu}(x) \to V(x) F_{\mu \nu}(x) V^{\dagger}(x)$ under a local gauge transformation $V(x)$, and is thus not gauge invariant on its own, unlike its counterpart in Abelian gauge theories. As we will show, the difficulty of using axial gauge is deeply connected with the configuration of the Wilson lines. We will also illuminate under what conditions a naive application of axial gauge leads to a correct result. 

\textbf{The axial gauge puzzle in the QGP.} We first discuss the puzzle in more detail by taking the example of the QGP transport coefficients for heavy quarks and quarkonia. The heavy quark diffusion coefficient is defined in terms of the zero frequency limit of a chromoelectric field correlator~\cite{Casalderrey-Solana:2006fio,CaronHuot:2007gq}
\begin{align}
g_E^{\rm Q}(t) = g^2 \left\langle {\rm Tr}_c \left( U_{[-\infty,t]} F_{0i}(t) U_{[t,0]} F_{0i}(0) U_{[0,-\infty]} \right) \right\rangle \label{HQ-corr} \,,
\end{align}
where $g$ denotes the strong coupling, angular brackets represent a thermal expectation value $\langle O \rangle \equiv {\rm Tr}(O \rho)$ with $\rho = e^{-\beta H}/{\rm Tr}(e^{-\beta H})$ and all fields are evaluated at the same spatial point, which is dropped here for notational simplicity. The field operators are ordered as shown (similarly below). The Wilson line $U_{[x,y]}$ is defined in the fundamental representation
\be
U_{[x,y]} = {\rm P} \exp \left( ig \int_y^x \!\! \diff z^\mu A_\mu^a(z) T^a_F \right) \,,
\ee
where ${\rm P}$ denotes path ordering and the path is a straight line connecting the two ends.
The plasma property relevant for small-size quarkonium in-medium dynamics is encoded in a different chromoelectric field correlator in both the quantum optical~\cite{Yao:2020eqy} and quantum Brownian motion limits~\cite{Brambilla:2016wgg,Brambilla:2017zei} (see also Ref.~\cite{Yao:2021lus})
\begin{align}
g_E^{\rm Q\bar{Q}}(t) = g^2 T_F \big\langle  {F}_{0i}^a(t) \ml{W}_{[t, 0]}^{ab}
   {F}_{0i}^b(0)  \big\rangle \,, \label{QA-corr}
\end{align} 
where $\ml{W}_{[x,y]}$ is an adjoint straight Wilson line
\be
\ml{W}_{[x,y]} = {\rm P} \exp \left( ig \int_y^x \!\! \diff z^\mu A_\mu^a(z) T^a_A \right) \,,
\ee
with the adjoint generators $[T^a_A]^{bc} = -i f^{abc}$. Different notations for $g_E^{Q\bar{Q}}$ in the literature are unified in Supplemental Material (SM). The correlator for quarkonium was constructed by using the effective field theory potential nonrelativistic QCD~\cite{Brambilla:1999xf,Brambilla:2004jw} and the open quantum system framework~\cite{Akamatsu:2011se,Akamatsu:2014qsa,Katz:2015qja,Brambilla:2017zei,Blaizot:2017ypk,Kajimoto:2017rel,Blaizot:2018oev,Yao:2018nmy,Akamatsu:2018xim,Miura:2019ssi,Sharma:2019xum,Rothkopf:2019ipj,Akamatsu:2020ypb,Sharma:2021vvu,Yao:2021lus}.

\begin{figure}
\centering
\includegraphics[width=0.45\textwidth]{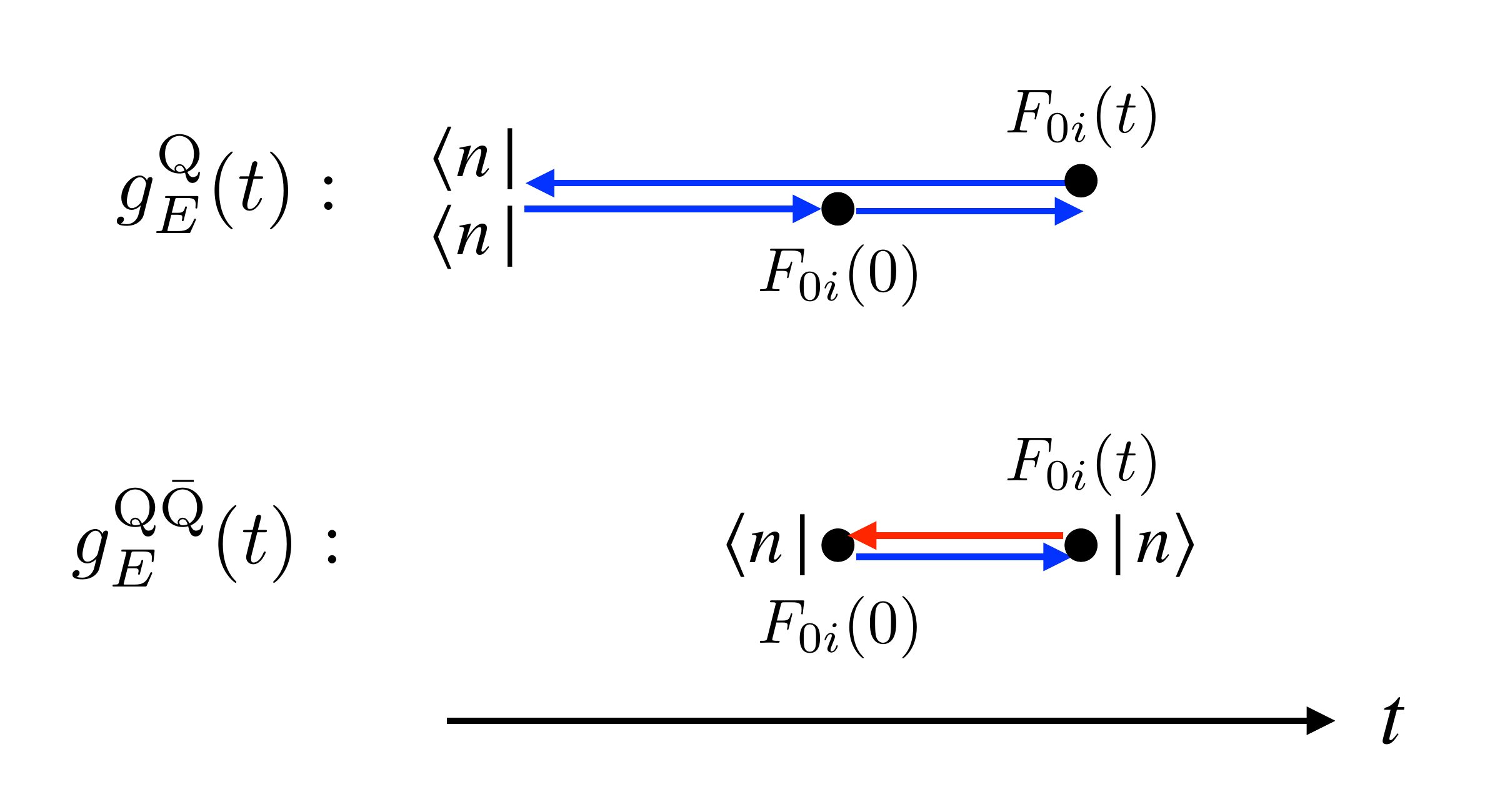}
\caption{Diagrammatic representation of the chromoelectric field correlators for open heavy quarks ($g_E^{\rm Q}(t)$, top row) and quarkonia ($g_E^{\rm Q\bar{Q}}(t)$, bottom row). The dots label the chromoelectric fields. The single and double lines with arrows indicate the Wilson lines in the fundamental and adjoint representations, respectively. The states $|n\rangle$ come from the trace $\Tr(O\rho) \propto \sum_n e^{-\beta E_n}\langle n| O |n \rangle$.}
\label{fig:correlator}
\end{figure}

These two correlators differ in their Wilson line configurations, as shown in Fig.~\ref{fig:correlator}, which contain important physical effects: The open heavy quark carries color through the diffusion process and the Wilson line accounts for both initial and final state interactions. For quarkonium, the Wilson line describes either initial or final state interaction~\cite{Yao:2021lus}. For quarkonium dissociation, the initial state is a heavy quark pair in color singlet which does not interact with the plasma at leading (zeroth) order in the multipole expansion while the final state is a pair in color octet, which does interact with the plasma at leading order, and vice versa for recombination. Explicit NLO calculations for $p_0>0$ showed that these two correlators are already different in vacuum (they also differ by temperature dependent terms, which we will not discuss here)~\cite{Burnier:2010rp,Eidemuller:1997bb,Eller:2019spw,Binder:2021otw}:
\begin{align}
\label{eq:diff-intro}
    \int_{-\infty}^{+\infty} \!\!\!\! \diff t \, e^{i p_0 t} \! \big(  g_E^{\rm Q\bar{Q}}(t) -  g_E^{\rm Q}(t) \big)_{\rm vac} \! = \frac{g^4 N_c (N_c^2 - 1) T_F p_0^3}{(2\pi)^3} \pi^2 \,.
\end{align}
However, these two correlators (\ref{HQ-corr}) and (\ref{QA-corr}) would become identical in temporal axial gauge where $A_0^a=0$ and all the Wilson lines become identities trivially. As a result, their difference is expected to vanish in axial gauge. Now we see the puzzle: The two correlators are defined gauge invariantly and calculations with different gauge choices should give the same result. However, the results in Feynman gauge and axial gauge are explicitly different.

\textbf{Resolution of the puzzle.} To resolve the puzzle, we first study the correlation functions in temporal axial gauge. For simplicity, we will only consider vacuum correlation functions, as their difference~\eqref{eq:diff-intro} is already apparent in vacuum. The time-ordered gluon propagator 
can be obtained using the FP procedure, which restricts the path integral over gauge field configurations to be on a ``slice'' determined by the gauge condition:\footnote{One can also choose $G^a[A]= G_A^a[A] - \omega^a(x)$ and then average over $\omega^a$, weighted by $\exp(-\frac{i}{2\xi}\int\diff^4x\, \omega^a\omega^a)$, where $\xi$ is a parameter. \label{fn:gauge-fixing} }
\be
G^a_A[A] = n^\mu A_\mu^a (x) \equiv 0 \,,
\ee
where throughout this section we will use $n^\mu=(1,0,0,0)$, i.e., temporal axial gauge. The time-ordered propagator is given by
\begin{align}
    &[D_T(k)]_{\mu \nu}^{ab} =  \\ & \frac{i \delta^{ab}}{k^2+i\varepsilon } \left[ -g_{\mu \nu} + \frac{n \! \cdot \! k \left( k_\mu n_\nu + n_\mu k_\nu \right) - n^2 k_\mu k_\nu}{(n \! \cdot \! k)^2 + i\varepsilon} \right] \, , \nn
\end{align}
where $\varepsilon \to 0$ is to be taken at the end of gauge-fixed calculations (we have dropped $\ml{O}(\varepsilon)$ terms in the numerator that do not contribute in this limit). The $\varepsilon$ prescription comes from the time-ordering prescription in the path integral (see SM). In temporal axial gauge, both correlation functions (\ref{HQ-corr}) and (\ref{QA-corr}) simply become $g^2 \langle 0| {\rm Tr}_c(E_i(t)E_i(0))|0\rangle$ for the vacuum part. An explicit NLO calculation for $p_0>0$ gives:\footnote{The Wightman and time-ordered correlators can be related by standard techniques.}
\begin{align}
    & \int_{-\infty}^{+\infty} \!\! \diff t\, e^{ip_0 t} \langle 0 |  g^2 \mathcal{T}( {E}^a_{i}(t)  {E}^a_{i}(0) ) | 0 \rangle = \\
    & \frac{g^2 (N_c^2 - 1) p_0^3}{(2\pi)^3}  \left\{ 4\pi^2 \! + \! N_c g^2 \! \left[ \frac{11}{12}\ln \! \left( \frac{\mu^2}{4p_0^2} \right) \! + \! \frac{149}{36} \! + \! \frac{\pi^2}{3} \right] \right\}\,,\nn
\end{align}
where $\ml{T}$ denotes time-ordering. This reproduces the Feynman gauge calculation result of Ref.~\cite{Eidemuller:1997bb} and matches the zero temperature limit of the result calculated in Ref.~\cite{Binder:2021otw} for Eq.~\eqref{QA-corr}. It also agrees with the corresponding Euclidean correlator in axial gauge (see SM).

The naive axial gauge calculation does not reproduce the Feynman gauge result for Eq.~\eqref{HQ-corr}, which implies that temporal axial gauge is not smoothly connected with Feynman gauge via a gauge transformation for this observable. To explicitly see the breakdown, we consider a more general gauge-fixing condition
\be
G_M^a[A] = \frac{1}{\lambda} n^\mu A_\mu^a (x) + \partial^\mu A^a_\mu(x) \, ,
\ee
which allows one to smoothly connect Feynman gauge (when $\lambda \to \infty$ for $\xi=1$)\footnote{The definition of $\xi$ is as in footnote~\ref{fn:gauge-fixing}, with $G_A^a$ replaced by $G_M^a$ in the gauge-fixing function.} with axial gauge (when $\lambda \to 0$ for any $\xi$). In this general gauge, the time-ordered gluon propagator with $\xi = 1$ becomes
\begin{align}
\label{eq:mixed-gauge}
    &[D_T(k)]_{\mu \nu}^{ab} = \frac{i \delta^{ab}}{k^2+i\varepsilon } \bigg[ -g_{\mu \nu} \\ 
    &  + \frac{  k_\mu n_\nu \left( n \! \cdot \! k - i \lambda k^2 \right) + n_\mu k_\nu \left( n \! \cdot \! k + i\lambda k^2 \right) - n^2 k_\mu k_\nu }{(n \! \cdot \! k)^2 + \lambda^2 (k^2)^2 + (1+ 2\lambda^2 k^2) i\varepsilon} \bigg] \,,\nn
\end{align}
where we have dropped $\ml{O}(\varepsilon)$ terms in the numerator. At any finite $\lambda$, one can evaluate the difference between Eqs.~\eqref{HQ-corr} and~\eqref{QA-corr} by carrying out the loop computations using Eq.~\eqref{eq:mixed-gauge}. We find the difference is the same as in Eq.~\eqref{eq:diff-intro} for any $\lambda \neq 0$ (see SM). Therefore, the difference between the two gauge invariant correlators is indeed preserved, even in the limit $\lambda \to 0$, as opposed to the conclusion one would have reached by naively setting $\lambda = 0$ from the start. The problem of naively setting $\lambda = 0$ is caused by a subtlety in the order of taking $\lambda \to 0$ and $\eta \to 0$, where $\eta$ is the regulator that implements how the Wilson line extends to infinity:
\be
U_{[(+\infty) n^\mu, 0]} = {\rm P} \exp \left( i g \int_0^{+\infty} \!\!\! \diff s \, e^{-\eta s} n^{\mu} A_\mu( s n^\mu) \right) \,.\qquad \label{eq:Wilson-line-regulated}
\ee
Specifically, in the calculation of the difference~\eqref{eq:diff-intro} 
there appear terms of the form (omitting color indices)
\begin{equation} \label{eq:problematic-terms}
\int \frac{\diff^4k}{(2\pi)^4} \frac{\eta}{ (n \! \cdot \! k)^2 + \eta^2}  \left[ D_T(k) \right]_{\nu \mu}  n^{\mu} N(p,k) \, ,
\end{equation}
which are sensitive to the order in which limits are taken. Here $N(p,k)$ is some function of external momentum $p$ and loop momentum $k$ that does not have poles at $ n \cdot k =0 $ or at $k^2=0$. If $\lambda \to 0$ is taken first, then $\lim_{\lambda \to 0} \left[ D_T(k) \right]_{\nu \mu}  n^\mu = 0$, and the result is zero. On the other hand, if one takes $\eta \to 0$ first, 
Eq.~\eqref{eq:problematic-terms} becomes
\begin{equation}
\int \frac{\diff^4k}{(2\pi)^4} \pi \delta( n \! \cdot \! k ) \left[ D_T(k) \right]_{\nu \mu} n^{\mu} N(p,k) \, ,
\end{equation}
and using the delta function leads to
\begin{align}
&\left. \left[ D_T(k) \right]_{\nu \mu} n^{\mu} \right|_{n \cdot k = 0 } \\
&\qquad\qquad  = \frac{(-1)}{k^2 + i\varepsilon} \left[ i n_{\nu} + \frac{  \lambda k^2 k_\nu }{ (\lambda k^2)^2 + (1 + 2\lambda^2 k^2) i\varepsilon } \right] \, , \nn
\end{align}
where we have used $n^2 = 1$.
We find the first term (proportional to $n_{\nu}$) gives the corresponding Feynman gauge result, and the second gives a vanishing contribution for all $\lambda$ (see SM). Thus, in one order of limits Eq.~\eqref{eq:problematic-terms} is trivially vanishing, whereas in the other it agrees with the Feynman gauge result.
Therefore, we conclude that the gauge invariant result can be reproduced in this general gauge, and naively imposing axial gauge leads to an incorrect result due to an order-of-limit subtlety. 
The correct order for calculating any physical observable that involves Wilson lines extending to infinity is to take $\eta \to 0$ first (as this defines the Wilson line), and then vary $\lambda$.\footnote{Our discussion here is about real-time quantities. However, we note that the imaginary time counterparts of Eqs.~\eqref{HQ-corr} and~\eqref{QA-corr} at finite temperature only involve Wilson lines of finite extent. In this case, one cannot gauge-fix the path integral to calculate Eq.~\eqref{HQ-corr} in temporal axial gauge because the Wilson line wraps around the periodic Euclidean time direction, which obtains contributions from gauge field configurations with nontrivial holonomy. In particular, the Polyakov loop would be trivial if such a gauge transformation were possible. However, the Polyakov loop is nontrivial and contains a wealth of information about QCD~\cite{Weiss:1980rj,Gross:1980br}.}

\textbf{A nonperturbative viewpoint.} Another way of seeing the problem in naive axial gauge is to scrutinize the path integral. The FP path integral of a pure gauge theory can be written as
\begin{align}
\int \ml{D}A\, {\rm det}\Big( \frac{\delta G^a(x)}{\delta\theta^b(y)} \Big) \prod_{x,a}\delta \big( G^a(x) \big) e^{iS_{\rm YM}[A^a]} \,,
\end{align} 
where $\theta^b(y)$ denotes the parameter specifying a gauge transformation, $G^a(x)$ is some gauge-fixing condition and $S_{\rm YM}[A^a]$ stands for the action of gauge fields.

We first illustrate the problem in the Abelian case by constructing a gauge transformation that connects Feynman gauge with axial gauge. Under a gauge transformation specified by $\theta(x)$, the Feynman gauge condition transforms as
\begin{align}
G_F(x): \quad \partial_\mu A^\mu(x) \to \partial_\mu A^\mu(x) - \partial^2 \theta(x) \,.
\end{align}
Setting $\partial_\mu A^\mu(x) - \partial^2 \theta(x) = G_A(x) = n_\mu A^\mu(x)$ in momentum space leads to
\begin{align}
& -i k_\mu A^\mu(k) + k^2 \theta(k) = n_\mu A^\mu(k) \,, \nonumber \\
&\implies \theta(k) = \frac{1}{k^2}\big( n_\mu A^\mu(k) + i k_\mu A^\mu(k) \big)\,.
\end{align}
Thus, the gauge transformation needed to transform Feynman gauge to axial gauge is given by
\begin{align}
A^\mu(k) \to M^\mu_{\ \nu} A^\nu(k) \,,
\end{align}
where the transformation matrix is set as
\begin{align}
M^\mu_{\ \nu} = g^\mu_{\ \nu} + \frac{ik^\mu}{k^2}\big( n_\nu + ik_\nu \big) \,.
\end{align}
Inspecting this matrix, we find that $k^\mu$ is an eigenvector of the transformation matrix $M$
\begin{align}
M^\mu_{\ \nu} k^\nu = i\frac{ n \cdot k}{k^2} k^\mu \,,
\end{align}
with eigenvalue $i (n \cdot k)/k^2$. Now we see that the gauge transformation is ill-defined when $n \cdot k = 0$ because the Jacobian of the transformation ${\rm det}(M)$ vanishes, which means the $n\cdot k = 0$ Fourier modes cannot be transformed in this way. Therefore, one cannot transform $n^\mu A_\mu(n \cdot  k = 0)$ to axial gauge from Feynman gauge, that generically has $n^\mu A_\mu(n \cdot  k = 0)\neq0$.

The nonexistence of such a gauge transformation does not always result in the breakdown of axial gauge calculations. To see this more clearly, we consider the gauge field at an arbitrary ``time'' $\bar{n} \cdot x$ ($\bar{n}$ is defined as $(1,-{\bs n})/\sqrt{1+{\bs n}^2}$ for $n=(1,{\bs n})/\sqrt{1+{\bs n}^2}$ so the coordinate $\bar{n} \cdot x$ is the Fourier conjugate of the momentum $n\cdot k$)
\be
A(\bar{n} \cdot x) = \int \diff(n\cdot k)\, e^{i (\bar{n} \cdot x)( n\cdot k) } A(n \cdot k) \,.
\ee
For finite $\bar{n} \cdot x$, the contribution at the point $n\cdot k = 0 $ can typically be neglected, since it has zero measure. However, at infinite ``time'' $\bar{n} \cdot x=\infty$, the dominant contribution to the Fourier transform comes from the region $n \cdot k \approx 0 $. Therefore, when gauge fields at infinite ``time'' are involved in calculations of correlation functions, the breakdown of the gauge transformation at $n \cdot k = 0 $ prevents us from properly gauge-fixing the path integral to axial gauge $n^\mu A_\mu = 0$ globally to perform the calculations. On the other hand, if the correlation function contains no gauge fields at infinite ``time'', the breakdown of the gauge transformation at $n \cdot k = 0$ is irrelevant, since the path integral of fields at $n \cdot k = 0$ only contributes to an overall normalization.
In the latter case, axial gauge calculations work well and give the correct result.

In the non-Abelian case the same breakdown can be seen even more simply by looking at the set of possible local gauge transformations acting on $A_\mu = A_\mu^a T_F^a$:
\begin{align}
A_{\mu}'(x) = V(x) A_\mu(x) V^{-1}(x) - \frac{i}{g} \big(\partial_\mu V(x)\big) V^{-1}(x) \, ,
\end{align}
where $V(x) = \exp(i \alpha^a(x) T_F^a)$. 
For the first term to be well-defined as $\bar{n} \! \cdot \! x \to \infty$, it is necessary that $\lim_{\bar{n}  \cdot  x \to  \infty} \alpha^a(x)$ exists,
which means the $\alpha^a(x)$ specifying the gauge transformation satisfies
\be
\lim_{\bar{n} \cdot x \to \infty} n^\mu \partial_{\mu} \alpha^a(x) = 0 \,.
\ee
Thus, when $\bar{n} \cdot x\! \to \infty$ the projection of the gauge field onto $n^\mu$ only transforms with an SU$(N_c)$ rotation
\begin{equation}
\left. n^{\mu} A_{\mu}' \right|_{\bar{n} \cdot x \to \infty} = \left. V n^\mu A_{\mu} V^{-1} \right|_{\bar{n} \cdot x \to \infty} \, ,
\end{equation}
with no ``shift'' term.
In particular, this means that ${\rm Tr}[ (n^\mu A_\mu (\bar{n} \cdot x = \infty))^2 ]$ cannot be changed by any gauge transformation. Therefore, if we start with a gauge choice in which $n^{\mu} A_\mu^a(\bar{n} \cdot x = \infty) \neq 0 $, we will not be able to set axial gauge $n^{\mu} A_\mu^a(\bar{n} \cdot x = \infty) = 0 $ via gauge transformations.
If the expectation value of an observable ${O}$ has finite contributions from gauge field configurations with $n^{\mu} A_\mu^a(\bar{n} \cdot x = \infty) \neq 0$ in a gauge-fixed path integral, or equivalently from the $n  \cdot  k = 0$ mode of $n^\mu A_\mu$, these contributions cannot be gauge-transformed away. Moreover, the corresponding axial gauge condition $n^\mu A_\mu^a = 0$ becomes inadequate. This is the case for the correlator that defines the heavy quark diffusion coefficient~\eqref{HQ-corr}. On the other hand, when the expectation value of an observable does not contain contributions from the field $n^{\mu} A_\mu^a(\bar{n} \cdot x = \infty)$ in the path integral, it is possible to operationally set $n^{\mu} A_\mu^a = 0$ everywhere and use axial gauge naively. However, we stress that this is possible not because one can effectively set axial gauge for all spacetime points in the path integral, but because the fields at the spacetime points where one cannot do so have no contributions to the expectation value of the operator. This is the case for the quarkonium correlator~\eqref{QA-corr}.

\textbf{Implications for other physical observables.} Finally, we discuss the implications of our findings on field strength correlators in other physical contexts. 
In the studies of TMDs, two gluon distributions with different Wilson line configurations exist. The Weizsaecker-Williams (WW) gluon TMD is defined by~\cite{Collins:1981uw,Mulders:2000sh,Ji:2005nu,Meissner:2007rx}
\begin{align}
& \frac{1}{xP^+} \int \frac{\diff b^- \diff b^2_\perp}{2(2\pi)^3} e^{-ixb^-P^+ -ib_
\perp \cdot k_\perp }  \\
& T_F \big\langle p(P,S) \big|  F^{a+i}(b^-,b_\perp) \ml{W}^{ad} F^{d+j}(0^-,0_\perp) \big| p(P,S) \big\rangle \,, \nn
\end{align}
where $|p(P,S)\rangle$ denotes the proton state with momentum $P$ and spin $S$. The adjoint Wilson line is $\ml{W}^{ad}=\ml{W}^{ab}_{[(b^-,b_\perp), (+\infty^-,b_\perp)]}\times \ml{W}^{bc}_{[(+\infty^-,b_\perp), (+\infty^-,0_\perp)]}\times \ml{W}^{cd}_{[(+\infty^-,0_\perp), (0^-,0_\perp)]}$.
The dipole gluon TMD is defined as~\cite{Kharzeev:2003wz,Dominguez:2010xd}
\begin{align}
& \frac{1}{xP^+} \int \frac{\diff b^- \diff b^2_\perp}{2(2\pi)^3} e^{-ixb^-P^+ -ib_
\perp \cdot k_\perp }  \\
&  \big\langle p(P,S) \big| {\rm Tr}_c \big[ U_1F^{+i}(b^-,b_\perp)  U_2
 F^{+j}(0^-,0_\perp) U_3 \big]
\big| p(P,S) \big\rangle  \,, \nn
\end{align}
where $U_1=U_{[(-\infty^-,0_\perp), (-\infty^-,b_\perp)]}  U_{[(-\infty^-,b_\perp), (b^-,b_\perp)]}$, $U_2=U_{[(b^-,b_\perp), (+\infty^-,b_\perp)]}\times U_{[(+\infty^-,b_\perp), (+\infty^-,0_\perp)]}\times U_{[(+\infty^-,0_\perp), (0^-,0_\perp)]}$, and $U_3=U_{[(0^-,0_\perp), (-\infty^-,0_\perp)]}$ are fundamental Wilson lines.
Their difference is well known~\cite{Kharzeev:2003wz,Dominguez:2010xd,Dominguez:2011wm,Metz:2011wb,Albacete:2012xq,Dumitru:2015gaa,Yao:2018vcg} from small-$x$ studies using the Color Glass Condensate framework~\cite{Iancu:2002xk,Jalilian-Marian:2005ccm,Gelis:2010nm}. They have different $k_\perp$ dependence for small $k_\perp$ while the high $k_\perp$ behavior is the same $\sim1/k_\perp^2$. Therefore, after integrating over the transverse momentum $k_\perp$ and averaging over the spins, we have two different gluon distributions
\begin{align}
\label{eqn:WWpdf}
&\frac{1}{xP^+} \int \frac{\diff b^- }{2(2\pi)} e^{-ixb^-P^+ } \\
& \qquad T_F \big \langle p(P) \big| F^{a+i}(b^-) \ml{W}^{ab}_{[b^-,0^-]} F^{b+j}(0^-) \big| p(P) \big\rangle \,,\nn\\
\label{eqn:dipolepdf}
&\frac{1}{xP^+} \int \frac{\diff b^- }{2(2\pi)} e^{-ixb^-P^+ } \big \langle p(P) \big| {\rm Tr}_c \big[ U_{[-\infty^-,b^-]} \\
& \qquad F^{+i}(b^-) U_{[b^-,0^-]}  F^{+j}(0^-) U_{[0^-,-\infty^-]} \big] 
\big| p(P) \big\rangle \, .\nn
\end{align}
Naively one would use~\cite{Bomhof:2006dp}
\begin{align}
\label{eqn:FtoA}
U_{[-\infty^-,0^-]} F^{a}_{\mu\nu}(0)T_F^a  U_{[0^-,-\infty^-]} = T^a_F \ml{W}_{[-\infty^-,0^-]}^{ab} F^{b}_{\mu\nu}(0) \,,
\end{align}
to show the two integrated gluon parton distribution functions (PDF) were the same. But Eq.~\eqref{eqn:FtoA} is only valid classically. In quantum theory, Eq.~\eqref{eqn:FtoA} only holds if a path ordering is applied on the left hand side for a path from $0^-$ to $-\infty^-$. Furthermore, the traditional wisdom that inserting a time-ordering operator does not change the physical meaning of the quark PDF~\cite{Jaffe:1983hp} may only apply for Eq.~\eqref{eqn:WWpdf} but not for Eq.~\eqref{eqn:dipolepdf} since the argument given therein relies on using light-cone gauge, which cannot be used naively for Eq.~\eqref{eqn:dipolepdf}.
Thus, these two integrated unpolarized gluon PDFs differ in terms of the operator orderings, similarly to the heavy quark~\eqref{HQ-corr} and quarkonium~\eqref{QA-corr} correlators. 
Therefore, our findings indicate that even though their expressions are identical in naive light-cone gauge, they may have different values. Future work should investigate whether time-ordering can be inserted into Eq.~\eqref{eqn:dipolepdf} without changing its meaning and whether the two gluon PDFs have the same value. If not, gluon PDFs are process dependent and their experimental determination needs systematic reanalysing. 

It is also important to discuss in what situations axial gauges still give physically sensible results, even when the observable involves Wilson lines of infinite extent. One ubiquitous such situation is when the observable is constructed as a limit of gauge invariant quantities of finite extent. 
We illustrate this by taking the jet quenching parameter as an example. Albeit not a field strength correlator, it is defined in terms of a Wilson loop~\cite{Wiedemann:2000za,Kovner:2003zj,Liu:2006ug,DEramo:2010wup,DEramo:2012uzl}:
\begin{align}
\hat{q} = \lim_{L^-\to+\infty^-} \frac{\sqrt{2}}{L^-}\int\frac{\diff^2k_\perp}{(2\pi)^2} k_\perp^2 \int\diff^2x_\perp e^{-ix_\perp\cdot k_\perp} \frac{\langle {\rm Tr}_c W^{\ml{R}}_\square \rangle}{d_{\ml{R}}}  \,, \label{eq:jet-quenching}
\end{align}
where $L^-$ is the length of the QGP medium along the $n^\mu=(1,0,0,1)/\sqrt{2}$ direction and $W^{\ml{R}}_\square$ is a rectangular Wilson loop with four corners at $(0^-,0_\perp)$, $(L^-,0_\perp)$, $(L^-,x_\perp)$ and $(0^-,x_\perp)$ in representation $\ml{R}$ of dimension $d_{\ml{R}}$. This is a manifestly gauge invariant object. In light-cone gauge, the transverse gauge links at fixed $0^-$ and $L^-$ become essential, along which $\diff z^\mu A_\mu \neq 0$. Furthermore, the introduction of these transverse gauge links enforces the Wilson loop to lie entirely within a finite spacetime volume in practical calculations, since the transverse gauge links can only be calculated at a finite light-cone distance $L^-$. Once this Wilson loop is calculated, one can take $L^- \to +\infty^-$ and obtain a physically sensible result (even in light-cone gauge). This is also why the light-cone gauge calculation of Ref.~\cite{Belitsky:2002sm} works for TMDs.

\textbf{Summary and Conclusions.} To summarize, in this letter we have scrutinized the applicability of axial gauge for computing gauge invariant non-Abelian field strength correlators. We found that attempting to gauge-fix the theory to axial gauge $n^{\mu} A_\mu = 0$
runs into an obstruction for the fields with the Fourier mode $n \cdot k = 0$. 
As a result, naive axial gauge is not reliable to remove $n^{\mu} A_\mu$ terms when one calculates correlators containing gauge fields at infinite ``time'' $\bar{n} \cdot x$, but it works well for correlators containing only gauge fields at finite $\bar{n} \cdot x$. Our studies further verify the difference between the two correlators defining the heavy quark and quarkonium transport coefficients, which means it is unjustified to use the heavy quark diffusion coefficient calculated via lattice field theory methods~\cite{Banerjee:2011ra,Francis:2015daa,Brambilla:2020siz,Altenkort:2020fgs} in quantum transport equations for small-size quarkonia, as done in Refs.~\cite{Brambilla:2020qwo,Brambilla:2021wkt}. Their difference beyond NLO is unknown. Therefore it is important to study their difference nonperturbatively via lattice field theory methods or the AdS/CFT correspondence~\cite{Maldacena:1997re,casalderrey2014gauge}. (The correlator~\eqref{HQ-corr} has been calculated at strong coupling using the AdS/CFT technique~\cite{Casalderrey-Solana:2006fio}.) Our findings also provide insights into the difference between the unpolarized WW and dipole gluon PDFs, which should be further investigated in the future. These studies will deepen our understanding of QGP transport properties and hadronic structure.

\begin{acknowledgments}
We are grateful for useful comments from Nora Brambilla, Adrian Dumitru and Krishna Rajagopal. We also want to thank Bob Jaffe for inspiring discussions on inserting a time-ordering operator in the definition of the quark PDF.
This work is supported by the U.S. Department of Energy, Office of Science, Office of Nuclear Physics under Grant Contract No. DE-SC0011090. XY is also supported in part by the U.S. Department of Energy, Office of Science, Office of Nuclear Physics, Inqubator for Quantum Simulation (IQuS) under Award Number DOE (NP) Award DE-SC0020970.
\end{acknowledgments}

\bibliography{main.bib}
\begin{widetext}
\newpage
\section{Supplemental Material}

\subsection{Different expressions for the quarkonium correlator present in the literature}

In this section we review the different definitions present in the literature~\cite{Brambilla:2016wgg,Brambilla:2017zei,Yao:2020eqy} that appear in the quantum transport equations for quarkonium. The following discussion will further illuminate the nature of the difference between the correlators~\eqref{HQ-corr} and~\eqref{QA-corr}.

We start from the definition of the chromoelectric correlator for the quarkonium transport equation as given in the main text~\eqref{QA-corr}, which appears in the formulation of the open quantum system for quarkonium in the quantum optical limit~\cite{Yao:2020eqy}:
\be
\label{app:QA}
g_E^{\rm Q\bar{Q}}(t) = g^2 T_F \left\langle  {F}_{0i}^a(t) \ml{W}_{[t, 0]}^{ab}
   {F}_{0i}^b(0) \right\rangle \, .
\ee
In the quantum optical limit, it is the correlator~\eqref{app:QA} at finite frequency that contributes to the quarkonium dissociation and recombination rates:
\be
g_E^{Q\bar{Q}}(p_0) = g^2 T_F \int_{-\infty}^{+\infty} \!\! \diff t \, e^{ip_0 t} \left\langle  {F}_{0i}^a(t) \ml{W}_{[t, 0]}^{ab} {F}_{0i}^b(0) \right\rangle \,.
\ee
On the other hand, in the quantum Brownian motion limit studied in Refs.~\cite{Brambilla:2016wgg,Brambilla:2017zei}, it is the zero frequency of $g_E^{Q\bar{Q}}$ that matters in the quarkonium transport (see also Ref.~\cite{Yao:2021lus}). However, the expression given in Refs.~\cite{Brambilla:2016wgg,Brambilla:2017zei} (see also Eq.~(2.13) of Ref.~\cite{Brambilla:2020qwo}) is
\begin{align}
\kappa^{Q\bar{Q}} \equiv \frac{g^2T_F}{N_c} {\rm Re} \int_{-\infty}^{+\infty}\!\!\diff t \, \Big\langle \ml{T}\Big( \widetilde{E}_i^a(t) \widetilde{E}_i^a(0) \Big) \Big\rangle\,,
\end{align}
where $\ml{T}$ denotes time-ordering and $\widetilde{E}_i^a(t) = U_{[-\infty,t]} E_i^a(t) U_{[t,-\infty]} = U_{[-\infty,t]} F_{0i}^a(t) U_{[t,-\infty]}$ with $U$ representing a fundamental Wilson line. The expression of $\kappa^{Q\bar{Q}}$ looks different from our expression $g_E^{Q\bar{Q}}(p_0=0)$ here. In the following, we will show they are equivalent for $p_0=0$. First we find
\begin{align}
\int_{-\infty}^{+\infty} \!\! \diff t \left\langle \ml{T} \left( {F}_{0i}^a(t) \ml{W}_{[t, 0]}^{ab} {F}_{0i}^b(0) \right) \right\rangle
& = \int_{0}^{+\infty} \!\! \diff t \left\langle  {F}_{0i}^a(t) \ml{W}_{[t, 0]}^{ab} {F}_{0i}^b(0) \right\rangle + \int_{-\infty}^{0} \!\! \diff t \left\langle {F}_{0i}^a(0) \ml{W}_{[0, t]}^{ab}  {F}_{0i}^b(t) \right\rangle \\
& = \int_{0}^{+\infty} \!\! \diff t \left\langle  {F}_{0i}^a(t) \ml{W}_{[t, 0]}^{ab} {F}_{0i}^b(0)  \right\rangle + \int_{0}^{+\infty} \!\! \diff t \left\langle {F}_{0i}^a(0) \ml{W}_{[0, -t]}^{ab}  {F}_{0i}^b(-t)  \right\rangle \nn\\
& = 2 \int_{0}^{+\infty} \!\! \diff t \left\langle {F}_{0i}^a(t) \ml{W}_{[t, 0]}^{ab} {F}_{0i}^b(0) \right\rangle \,,\nn
\end{align} 
where we have relabeled the color indexes $a$ and $b$ in the second term on the first line, flipped the sign of $t$ in the second term on the second line and used translational invariance in $t$ on the last line. Then we can show
\begin{align}
{\rm Re} \int_{-\infty}^{+\infty} \!\! \diff t \left\langle \ml{T} \left( {F}_{0i}^a(t) \ml{W}_{[t, 0]}^{ab} {F}_{0i}^b(0) \right) \right\rangle & = 2\,{\rm Re} \int_{0}^{+\infty} \!\! \diff t \left\langle  {F}_{0i}^a(t) \ml{W}_{[t, 0]}^{ab} {F}_{0i}^b(0)  \right\rangle \\
& = \int_{0}^{+\infty} \!\! \diff t \left\langle  {F}_{0i}^a(t) \ml{W}_{[t, 0]}^{ab} {F}_{0i}^b(0)  \right\rangle + \int_{0}^{+\infty} \!\! \diff t \left\langle {F}_{0i}^a(0) \ml{W}_{[0, t]}^{ab} {F}_{0i}^b(t) \right\rangle \nn\\
& = \int_{0}^{+\infty} \!\! \diff t \left\langle {F}_{0i}^a(t) \ml{W}_{[t, 0]}^{ab} {F}_{0i}^b(0) \right\rangle + \int_{-\infty}^{0} \!\! \diff t \left\langle {F}_{0i}^a(0) \ml{W}_{[0, -t]}^{ab} {F}_{0i}^b(-t)  \right\rangle \nn\\
& = \int_{0}^{+\infty} \!\! \diff t \left\langle {F}_{0i}^a(t) \ml{W}_{[t, 0]}^{ab} {F}_{0i}^b(0)  \right\rangle + \int_{-\infty}^{0} \!\! \diff t \left\langle {F}_{0i}^a(t) \ml{W}_{[t, 0]}^{ab} {F}_{0i}^b(0)  \right\rangle \nn\\[4pt]
& \propto g_E^{Q\bar{Q}}(p_0=0) \,, \nn
\end{align}
where we have relabeled the color indexes $a$ and $b$ in the second term on the second line, flipped the sign of $t$ in the second term on the third line and used translational invariance in $t$ in the second term on the second-to-last line. Finally we study the relation between $\langle \ml{T} ( {F}_{0i}^a(t) \ml{W}_{[t, 0]}^{ab} {F}_{0i}^b(0) ) \rangle$ and $\langle \ml{T} ( \widetilde{E}_i^a(t) \widetilde{E}_i^a(0) ) \rangle$, both of which can be studied using the closed-time path integral methods (L. V. Keldysh et al., Sov. Phys. JETP 20, 1018 (1965)). Since both correlators are time-ordered, we can insert all the fields contained in the correlators on the time-ordered branch of the Schwinger-Keldysh contour, i.e., the upper branch with type-$1$ fields. Then we can use the standard SU$(N_c)$ Wilson line algebra to derive
\begin{align}
\label{eq:difference-correlators-explicit}
& g^2 T_F \left\langle \ml{T} \left( {F}_{0i}^a(t) \ml{W}_{[t, 0]}^{ab} {F}_{0i}^b(0) \right) \right\rangle \\
&= g^2 T_F \int \! DA_E DA_1 DA_2 \, e^{iS[A_1]-iS[A_2]-S_E[A_E]} F_{0i}^a[A_1](t) \, \ml{W}_{[t, 0]}^{ab}[A_1] \, F_{0i}^b[A_1](0) \nonumber \\
    &= g^2 \int \! DA_E DA_1 DA_2 \, e^{iS[A_1]-iS[A_2]-S_E[A_E]} \, {\rm Tr}_{c} \! \left\{ F_{0i}[A_1](t) \, U_{[t, 0]}[A_1] \, F_{0i}[A_1](0) \,  U_{[0, t]}[A_1] \right\}  \nonumber \\
    &= g^2 \int \! DA_E DA_1 DA_2 \, e^{iS[A_1]-iS[A_2]-S_E[A_E]} \, {\rm Tr}_{c} \! \left\{ U_{[-\infty, t]}[A_1] \, F_{0i}[A_1](t) \, U_{[t, 0]}[A_1] \, F_{0i}[A_1](0) \, U_{[0, -\infty]}[A_1] \right\}  \nonumber \\
    &= g^2 \left\langle \mathcal{T} \, {\rm Tr}_c \left( U_{[-\infty,t]} F_{0i}(t) U_{[t,0]} F_{0i}(0) U_{[0,-\infty]} \right) \right\rangle \nonumber \\[4pt]
& = g^2T_F \langle \ml{T} ( \widetilde{E}_i^a(t) \widetilde{E}_i^a(0) ) \rangle \,,
 \nn
\end{align}
where the subscripts $1$, $2$ and $E$ denote the type-1, type-2 and Euclidean fields. 
Putting everything together, we have proved that $\kappa^{Q\bar{Q}}$ defined in Refs.~\cite{Brambilla:2016wgg,Brambilla:2017zei} and our expression $g_E^{Q\bar{Q}}(p_0=0)$ are the same, up to a trivial normalization factor.

We want to emphasize that the second-to-last line of this expression~\eqref{eq:difference-correlators-explicit} does not match the correlator that defines the heavy quark diffusion coefficient $\kappa^Q$~\cite{Casalderrey-Solana:2006fio}, which is given by
\be
\kappa^Q \propto g^2 {\rm Re} \int_{-\infty}^{+\infty} \!\!\diff t  \left\langle {\rm Tr}_c \left( U_{[-\infty,t]} F_{0i}(t) U_{[t,0]} F_{0i}(0) U_{[0,-\infty]} \right) \right\rangle  = 
{\rm Re} \int_{-\infty}^{+\infty} \!\!\diff t\, g_E^Q(t) \, .
\ee
The key difference between the second-to-last line of Eq.~\eqref{eq:difference-correlators-explicit} and $g_E^Q$ is the operator ordering: in the former case the operators are time-ordered while in the latter they are ordered in the sequence as shown. Conceptually they are different in the sense of Figure~\ref{fig:correlator}: the Wilson loop in~\eqref{HQ-corr} is interrupted by the (thermal) trace over states, whereas~\eqref{QA-corr} can be written in a way such that the Wilson lines only appear between the two chromoelectric field operators. In the original formulation of the heavy quark diffusion coefficient~\cite{Casalderrey-Solana:2006fio}, the Wilson line configuration wraps around the closed-time Schwinger-Keldysh contour with a winding number equal to one. The Euclidean calculation of the heavy quark diffusion coefficient~\cite{Burnier:2010rp} also has this feature (see~\cite{Eller:2019spw} for an explicit proof that the Minkowski formulation~\cite{Casalderrey-Solana:2006fio} and the Euclidean formulation~\cite{Burnier:2010rp} give the same result). The Wilson line configuration in the correlator for quarkonium has a winding number equal to zero.
This mathematical difference has physical origin as discussed in the main text. 
Therefore, these two quantities $\kappa^{Q\bar{Q}}$ and $\kappa^Q$ (or more generally $g_E^{\rm Q\bar{Q}}$ and $g_E^Q$) cannot be used interchangeably, which has also been noted in Ref.~\cite{Eller:2019spw}.

The first verification that $g_E^{\rm Q\bar{Q}}$ and $g_E^Q$ are different was achieved in~\cite{Burnier:2010rp}, which can be seen by comparing the results obtained therein for $g_E^Q$ to the results of~\cite{Eidemuller:1997bb}, which first computed the correlator $g_E^{\rm Q\bar{Q}}$ in vacuum. This difference was subsequently verified in~\cite{Binder:2021otw}, which calculates $g_E^{\rm Q\bar{Q}}$ both in vacuum and at finite temperature. Furthermore, the imaginary part of Eq.~\eqref{eq:difference-correlators-explicit} at zero frequency differs from the imaginary part of $g_E^{Q}$ at zero frequency already at $\ml{O}(g^4)$~\cite{Eller:2019spw}. In the present paper we further study their difference with a more general gauge choice, and show the breakdown of a naive axial gauge calculation.

\subsection{Chromoelectric correlator in axial gauge}
Here we study the chromoelectric field correlators for heavy quarks and quarkonia in axial gauge, in which the Wilson lines become identities and thus can be neglected. The time-ordered chromoelectric field correlator is 
\be
g_{E,\,T}^{\rm Axial}(p_0) = g^2 \int_{-\infty}^{+\infty} \! \diff t\, e^{i p_0 t} \langle 0 |  \mathcal{T} ( {E}^a_{i}(t,{\bs x})  {E}^a_{i}(0,{\bs x}) ) | 0 \rangle\,,
\ee
and we want to calculate it in axial gauge at next-to-leading order (NLO). We will focus on the gluon polarization diagram, which contributes to the correlator at NLO. First we work out the gluon propagator in axial gauge. The free part of the gauge boson Lagrangian plus the gauge-fixing term in momentum space can be written as
\begin{align}
\frac{i}{2} \int \diff^4 k \, A^{\mu a}(-k) \Big( - g_{\mu\nu}(k^2 +i\varepsilon) +k_\mu k_\nu - \frac{1}{\xi}n_\mu n_\nu \Big) A^{\nu a}(k) \,,
\end{align}
where $\varepsilon$ comes from the boundary condition of the path integral at $t=\pm\infty$ and $\xi$ is a gauge-fixing parameter to be set later. Inverting $i g_{\mu\nu}(k^2 +i\varepsilon) -i k_\mu k_\nu +i n_\mu n_\nu/\xi$ gives the time-ordered gluon propagator
\begin{align}
[D_T(k)]_{\mu\nu}^{ab} = \frac{i\delta^{ab}}{k^2+i\epsilon} \left[ -g_{\mu\nu} + \frac{ n\cdot k(k_\mu n_\nu + k_\nu n_\mu) - [\xi(k^2+i\varepsilon)+n^2] k_\mu k_\nu +i\varepsilon n_\mu n_\nu }{i\varepsilon[\xi(k^2+i\varepsilon)+n^2] + (n\cdot k)^2} \right] \,.
\end{align}
Setting $\xi=0$ and $n_{\mu}=(1,0,0,0)$ for temporal axial gauge and neglecting terms proportional to $\varepsilon$ in the numerator lead to
\begin{align}
[D_T(k)]_{\mu\nu}^{ab} = \frac{i\delta^{ab} P_{\mu\nu}(k) }{k^2+i\epsilon} \equiv D_T(k) \delta^{ab} P_{\mu\nu}(k) \,,
\end{align}
where
\begin{align}
    P_{\mu \nu}(k) = -g_{\mu \nu} + \frac{k_0(k_\mu n_\nu + n_\mu k_\nu)}{k_0^2 + i\varepsilon} - \frac{k_\mu k_\nu}{k_0^2 + i\varepsilon} \, .
\end{align}
Then, using the Feynman rules of non-Abelian gauge theory we find the contribution of the gluon polarization diagrams (with its two external legs connected with the two chromoelectric fields) to the time-ordered chromoelectric correlator
\begin{align}
\left. g_{E,\,T}^{\rm Axial}(p_0) \right|_{\rm NLO} &= g^2 \int_{\bs p} D_T(p)^2 (i p_0 g_i^{\,\sigma'} - i p_i g_0^{\,\sigma'}) P_{\sigma' \sigma}(p) (- i p_0 g_i^{\,\rho'} + i p_i g_0^{\,\rho'}) P_{\rho' \rho}(p) \nonumber \\
& \quad \frac12 \bigg( \delta^{cd} \int_k D_T(k) \delta^{ab} P_{\mu \nu}(k) (-i g^2) \left[ f^{abe} f^{cde} (g^{\mu \rho} g^{\nu \sigma} - g^{\mu \sigma} g^{\nu \rho}) + f^{ace} f^{dbe} (g^{\mu \sigma} g^{\nu \rho} - g^{\mu \nu} g^{\rho \sigma}  ) \right.  \nonumber \\
    & \quad \quad \quad \quad \quad \quad \quad \quad \quad \quad \quad \quad + \left. f^{ade} f^{bce} (g^{\mu \nu} g^{\rho \sigma} - g^{\mu \rho} g^{\nu \sigma}) \right] \nonumber \\
& + \delta^{cc'} \int_k D_T(k) D_T(p-k) \delta^{aa'} \delta^{bb'} P_{\mu\mu'}(k) P_{\nu\nu'}(p-k) g f^{abc} \left[ g^{\mu \nu} (p - 2k)^{\sigma} + g^{\nu \sigma} (k-2p)^{\mu} + g^{\sigma \mu}(p+k)^{\nu} \right] \nonumber \\
    & \quad \quad \quad \quad \quad \quad \quad \quad \times g f^{a'b'c'} \left[ g^{\mu' \nu'} (2k-p)^{\rho} + g^{\nu' \rho} (2p-k)^{\mu'} + g^{\rho \mu'}(-p-k)^{\nu'} \right] \bigg) \,,
\end{align}
where $\int_{\bs p} \equiv \int\frac{\diff^d p}{(2\pi)^d}$ and $\int_k \equiv \int\frac{\diff^D k}{(2\pi)^D}$.
Our strategy to evaluate these integrals is to do the ${\bs p}$ and ${\bs k}$ integrals first, using dimensional regularization in $d = 3 - \tilde{\epsilon}$ spatial dimensions for both of them. $D = 4 - \tilde{\epsilon}$ is the total number of spacetime dimensions. (The calculation is only consistent if we use the same dimensionality for both ${\bs p}$ and ${\bs k}$ integrals.) We leave the integral over $k_0$ to be done at the end of the calculation.
We proceed by reducing the integral into a handful of integral structures $\tilde{I}_i(p,k)$ and their respective numerators $N_i(p_0,k_0)$ that do not depend on spatial momenta, which gives
\be
g_{E,\,T}^{\rm Axial}(p_0) = \frac{N_c (N_c^2-1) g^4 p_0^2 }{2} \int_{-\infty}^{+\infty} \! \frac{\diff k_0} {2\pi} \int \frac{\diff^d {\bs p} \diff^d {\bs k} }{(2\pi)^{2d}} \sum_i N_i(p_0,k_0) \tilde{I}_i(p,k) \, .
\ee
Also, we denote $I_i(p_0,k_0) = \int \frac{\diff^d {\bs p} \diff^d {\bs k} }{(2\pi)^{2d}} \tilde{I}_i(p,k)$.

Below we list the resulting integral structures, accompanied by their respective numerators:
\begin{enumerate}
    \item \begin{align}
    I_1 &= \int_{\bs p} \int_{\bs k} \frac{1}{(p^2+i\varepsilon)^2 (k^2 + i\varepsilon) ((p-k)^2 + i\varepsilon) } \, , \\ 
    N_1 &= 0 \, , \end{align}
    \item \begin{align}
    I_2 &= \int_{\bs p} \int_{\bs k} \frac{1}{(p^2+i\varepsilon) (k^2 + i\varepsilon) ((p-k)^2 + i\varepsilon) } \nonumber \\ &= - \frac{\Gamma(4-D)}{(4\pi)^{D-1} } \int_0^1 \diff x \diff y \frac{ \left[ - y (k_0 - x p_0)^2 - (1 - y + y x(1-x)) p_0^2 - i\varepsilon \right]^{D-4} }{y^{\frac{D-3}{2}} (1-y+yx(1-x))^{\frac{D-1}{2}}} \, , \\ 
    N_2 &=  \frac{4 (D-2) (k_0^2 - k_0 p_0 + p_0^2)^2 }{k_0 (k_0 - p_0) p_0^2 } \, , \end{align}
    \item \begin{align}
    I_3 &= \int_{\bs p} \int_{\bs k} \frac{1}{(p^2+i\varepsilon)^2 (k^2 + i\varepsilon)} = - (-i)^{2D-8} \frac{\Gamma \! \left( \frac{3-D}{2} \right) \Gamma \! \left( \frac{5-D}{2} \right) }{(4\pi)^{D-1}} \frac{|k_0 p_0|^{D-3}}{p_0^2} \, , \\ 
    N_3 &=  \frac{2 (D-2) \left[ 2(D-1) k_0^2 - (D-2) p_0 k_0 + 2(D-1) p_0^2 \right] }{(D-1) k_0 p_0 } \, , \end{align}
    \item \begin{align}
    I_4 &= \int_{\bs p} \int_{\bs k} \frac{1}{(p^2+i\varepsilon) (k^2 + i\varepsilon)}  = (-i)^{2D-6} \frac{\Gamma\!\left(\frac{3-D}{2} \right)^2}{(4\pi)^{D-1}} |k_0 p_0|^{D-3} \, , \\ 
    N_4 &=  \frac{2 \left[ 2(D-2) k_0^4 - 3(D-2) k_0^3 p_0 + D\, k_0^2 p_0^2 - (D-2) k_0 p_0^3 + (D-2) p_0^4 \right]  }{ k_0^2 p_0^3 (k_0 - p_0) } \, , \end{align}
    \item \begin{align}
    I_5 &= \int_{\bs p} \int_{\bs k} \frac{1}{((p-k)^2+i\varepsilon) (k^2 + i\varepsilon)} = (-i)^{2D-6} \frac{\Gamma\!\left(\frac{3-D}{2} \right)^2}{(4\pi)^{D-1}} |k_0 (p_0-k_0) |^{D-3} \, , \\ 
    N_5 &=  \frac{D-2}{k_0^2} + \frac{D-2}{(k_0-p_0)^2} + \frac{2}{p_0^2} \, , \end{align}
\end{enumerate}
All the other integral structures give vanishing contributions in dimensional regularization. Note that the term 1.~vanishes because the numerator happens to be zero, and the term 3.~also vanishes upon integration over $k_0$ because the integrand is just a polynomial in $k_0$. (In dimensional regularization the limit $\tilde{\epsilon} \to 0$ cannot be taken before performing all integrals that involve $d$, which means the limit should be taken after the $k_0$ integral.) Then, one can show that for $p_0>0$
\begin{enumerate}
    \item \begin{align}
    \frac{N_c (N_c^2-1) g^4 p_0^2 }{2}  \int_{-\infty}^{+\infty} \frac{\diff k_0}{2\pi}  N_1(p_0,k_0) I_1(p,k) = 0
     \, , \end{align}
    \item \begin{align}
    \frac{N_c (N_c^2-1) g^4 p_0^2 }{2}  \int_{-\infty}^{+\infty} \frac{\diff k_0}{2\pi}  N_2(p_0,k_0) I_2(p,k) = \frac{N_c(N_c^2-1)g^4 p_0^3}{(2\pi)^3} \left[ \frac{11}{6 \tilde{\epsilon} } + \frac{11}{6} \ln \! \left( \frac{\mu^2}{4p_0^2} \right) + \frac{167}{36} + \frac{\pi^2}{3} \right]
     \, , \end{align}
    \item \begin{align}
    \frac{N_c (N_c^2-1) g^4 p_0^2 }{2} \int_{-\infty}^{+\infty} \frac{\diff k_0}{2\pi}  N_3(p_0,k_0) I_3(p,k) = 0
     \, , \end{align}
    \item \begin{align}
    \frac{N_c (N_c^2-1) g^4 p_0^2 }{2} \int_{-\infty}^{+\infty} \frac{\diff k_0}{2\pi}  \left[ N_4(p_0,k_0) I_4(p,k) + N_5(p_0,k_0) I_5(p_0,k_0) \right] = \frac{N_c (N_c^2 - 1) g^4 p_0^3}{ (2\pi)^3} \frac5{12}
     \, . \end{align}
\end{enumerate}

The final contribution to evaluate is from the coupling constant counterterm, since the definition of the chromoelectric correlator contains $g^2$. The contribution for $p_0>0$ reads
\begin{align}
    & (Z_g - 1 ) (N_c^2 - 1) g^2 \int_{\bs p}  D_T(p) (i p_0 g_i^{\,\sigma} - i p_i g_0^{\,\sigma}) P_{\sigma \sigma'}(p) (- i p_0 g_i^{\,\sigma'} + i p_i g_0^{\,\sigma'}) \nonumber \\
    &= \frac{g^4}{8\pi^2 (D-4)} \frac{11}{3} N_c (N_c^2 - 1) p_0^2 \int_{\bs p} \frac{i (\delta_{ii} - {\bs p}^2/p_0^2 )}{p_0^2 - {\bs p}^2 + i\varepsilon} \nonumber \\
    &= \frac{g^4 N_c (N_c^2 - 1)}{(2\pi)^3} p_0^2 \left[ \frac{11}{24 \pi^2 (D-4)} \frac{1}{(2\pi)^{D-4}} \pi \frac{1}{2p_0} \Omega_{D-1} ( D-2 ) p_0^{D-2}  \right] \nonumber \\
    &= \frac{g^4 N_c (N_c^2 - 1)}{(2\pi)^3} p_0^3 \frac{11}{48\pi} \left[ \frac{1}{ (D-4)} \frac{\Omega_{D-1} ( D-2 ) (p_0/\tilde{\mu} )^{D-4}}{(2\pi)^{D-4}}   \right] \nonumber \\
    &= \frac{g^4 N_c (N_c^2 - 1)}{(2\pi)^3} p_0^3 \left[ - \frac{11}{6\tilde{\epsilon}} + \frac{11}{48\pi} \frac{\partial}{\partial D} \left( \frac{\Omega_{D-1} ( D-2 ) (p_0/\tilde{\mu} )^{D-4}}{(2\pi)^{D-4}} \right)_{D=4} \right] \nonumber \\
    &= \frac{g^4 N_c (N_c^2 - 1)}{(2\pi)^3} p_0^3 \left[ - \frac{11}{6\tilde{\epsilon}} + \frac{11}{12} \left( 1 + \ln( \pi (p_0/(2\pi \tilde{\mu}))^2 ) - ( 2 - \gamma_E - 2\ln(2) ) \right) \right] \nonumber \\
    &=  \frac{g^4 N_c (N_c^2 - 1)}{(2\pi)^3} p_0^3 \left[ - \frac{11}{6\tilde{\epsilon}} - \frac{11}{12} - \frac{11}{12} \ln \! \left( \frac{\mu^2}{4p_0^2} \right)  \right] \,,
\end{align}
where $\mu^2 = 4\pi e^{-\gamma_E} \tilde{\mu}^2$.
Adding everything up, one obtains for $p_0>0$
\begin{equation}
    \left. g_{E,\,T}^{\rm Axial}(p_0) \right|_{\rm NLO}  = \frac{g^4 N_c (N_c^2 - 1)}{(2\pi)^3} p_0^3 \left[ \frac{11}{12}\ln \! \left( \frac{\mu^2}{4p_0^2} \right) + \frac{149}{36} + \frac{\pi^2}{3} \right],
\end{equation}
which is what we give in the main text.

The corresponding Euclidean correlator can be evaluated in exactly the same way. In the same notation (but with the understanding that $k_0$ and $p_0$ are now Euclidean quantities), the relevant integral structures are
\begin{enumerate}
    \item \begin{align}
    I_1 &= \int_{\bs p} \int_{\bs k} \frac{1}{(p^2)^2 k^2 (p-k)^2 } \, , \\ 
    N_1 &= 0 \, , \end{align}
    \item \begin{align}
    I_2 &= \int_{\bs p} \int_{\bs k} \frac{1}{p^2 k^2  (p-k)^2 } \nonumber \\ &=  \frac{\Gamma(4-D)}{(4\pi)^{D-1} } \int_0^1 \diff x \diff y \frac{ \left[  y (k_0 - x p_0)^2 + (1 - y + y x(1-x)) p_0^2 \right]^{D-4} }{y^{\frac{D-3}{2}} (1-y+yx(1-x))^{\frac{D-1}{2}}}  \, , \\ 
    N_2 &= - \frac{4 (D-2) (k_0^2 - k_0 p_0 + p_0^2)^2 }{k_0 (k_0 - p_0) p_0^2 } \, , \end{align}
    \item \begin{align}
    I_3 &= \int_{\bs p} \int_{\bs k} \frac{1}{(p^2)^2 k^2} = \frac{\Gamma \! \left( \frac{3-D}{2} \right) \Gamma \! \left( \frac{5-D}{2} \right) }{(4\pi)^{D-1}} \frac{|k_0 p_0|^{D-3}}{p_0^2} \, , \\ 
    N_3 &= - \frac{2 (D-2) \left[ 2(D-1) k_0^2 - (D-2) p_0 k_0 + 2(D-1) p_0^2 \right] }{(D-1) k_0 p_0 } \, , \end{align}
    \item \begin{align}
    I_4 &= \int_{\bs p} \int_{\bs k} \frac{1}{p^2 k^2 }  =  \frac{\Gamma\!\left(\frac{3-D}{2} \right)^2}{(4\pi)^{D-1}} |k_0 p_0|^{D-3} \, , \\ 
    N_4 &=  -\frac{2 \left[ 2(D-2) k_0^4 - 3(D-2) k_0^3 p_0 + D k_0^2 p_0^2 - (D-2) k_0 p_0^3 + (D-2) p_0^4 \right]  }{ k_0^2 p_0^3 (k_0 - p_0) } \, , \end{align}
    \item \begin{align}
    I_5 &= \int_{\bs p} \int_{\bs k} \frac{1}{(p-k)^2 k^2 } = \frac{\Gamma\!\left(\frac{3-D}{2} \right)^2}{(4\pi)^{D-1}} |k_0 (p_0-k_0) |^{D-3} \, , \\ 
    N_5 &= - \frac{D-2}{k_0^2} - \frac{D-2}{(k_0-p_0)^2} - \frac{2}{p_0^2} \, , \end{align}
\end{enumerate}
all of which give the same contributions to the correlator as in the time-ordered case.

\subsection{Difference between heavy quark and quarkonium correlators in mixed axial-Feynman gauge}

In this section we show how to calculate the difference between the two correlators (heavy quark and quarkonium) discussed in the main text in mixed axial-Feynman gauge. To enforce the different operator orderings in the correlation functions, a rigorous calculation of their difference (in particular, the heavy quark correlator) requires introducing the Schwinger-Keldysh contour (L. V. Keldysh et al., Sov. Phys. JETP 20, 1018 (1965)). As mentioned in the main text, we use a mixed gauge-fixing condition given by
\be
G_M^a[A] = \frac{1}{\lambda} n^\mu A_\mu^a (x) + \partial^\mu A^a_\mu(x) \, ,
\ee
over which we perform the standard average over field configurations $\delta(G_M^a[A] - \omega^a(x))$ weighted by a function $\exp(-\frac{i}{2\xi}\int\diff^4x \omega^a\omega^a)$ with a smearing parameter $\xi$~\cite{Srednicki:2007qs,Schwartz:2014sze}. When $\lambda \to \infty$, one recovers Feynman gauge by setting $\xi = 1$ while axial gauge is recovered for $\lambda \to 0$ for any $\xi$. As such, we shall choose $\xi = 1$ throughout. After performing the gauge-fixing procedure in the path integral, the gluon propagators in this gauge are obtained by inverting the kinetic term in the action for the gauge field: 
\begin{align}
    &i S_{\rm kin}[A] = \\
    & -\frac12 \int_k \begin{pmatrix} \\ A_{(1)}^{a \nu}(-k) \\ \\ A_{(2)}^{a \nu}(-k) \\ {} \end{pmatrix}^{T} \begin{pmatrix} \begin{matrix} i \Big[ \left(k^2 + i\varepsilon \right) g_{\mu \nu} - \left(1 - \frac{1}{\xi} \right) k_\mu k_\nu \\ \,\,\, + \frac{1}{\xi \lambda^2} n_\mu n_{\nu} + \frac{1}{\xi \lambda} \left( - i k_\mu n_\nu + i k_\nu n_\mu \right)  \Big] \end{matrix}  & 2\varepsilon g_{\mu \nu} \Theta(-k_0) \\ 2\varepsilon g_{\mu \nu} \Theta(k_0) &  \begin{matrix} -i \Big[ \left(k^2 - i\varepsilon \right) g_{\mu \nu} - \left(1 - \frac{1}{\xi} \right) k_\mu k_\nu \\ \,\,\, + \frac{1}{\xi \lambda^2} n_\mu n_{\nu} + \frac{1}{\xi \lambda} \left( - i k_\mu n_\nu + i k_\nu n_\mu \right)  \Big] \end{matrix} \end{pmatrix} \begin{pmatrix} \\ A_{(1)}^{a \mu}(k) \\ \\ A_{(2)}^{a \mu}(k) \\ {} \end{pmatrix} \nn
\end{align}
and we will set $\xi = 1$ in what follows.
The resulting propagators read
\begin{align}
    [D_T(k)]_{\mu\nu}^{ab} &= \frac{i\delta^{ab}}{k^2+i\varepsilon} \left[ -g_{\mu\nu} + \frac{k_\mu n_\nu f + k_\nu n_\mu f^* - n^2 k_\mu k_\nu }{ |f|^2 + i h \varepsilon } \right] \,, \\
    [D_{\bar{T}}(k)]_{\mu \nu}^{ab} &= \frac{-i\delta^{ab}}{k^2-i\varepsilon} \left[ -g_{\mu\nu} + \frac{k_\mu n_\nu f + k_\nu n_\mu f^* - n^2 k_\mu k_\nu }{ |f|^2 - i h \varepsilon } \right] \,, \\
    [D_>(k)]_{\mu\nu}^{ab} &= \frac{2\varepsilon \Theta(k_0) }{(k^2)^2 + \varepsilon^2} \left[ - g_{\mu \nu} + \frac{k_\mu n_\nu F + k_\nu n_\mu F^* -  k_\mu k_\nu H }{|f|^4 + h^2 \varepsilon^2 } \right] \,, \\
    [D_<(k)]_{\mu\nu}^{ab} &= \frac{2\varepsilon \Theta(-k_0) }{(k^2)^2 + \varepsilon^2} \left[ - g_{\mu \nu} + \frac{k_\mu n_\nu F + k_\nu n_\mu F^* -  k_\mu k_\nu H }{|f|^4 + h^2 \varepsilon^2 } \right] \,,
\end{align}
where we have denoted for brevity
\begin{align}
    f &= (n \! \cdot \! k - i\lambda k^2) \,,  \\
    h &= (n^2+ 2\lambda^2 k^2) \,, \\ 
    F &= k^2 n^2 f + 2 f |f|^2 -  (n \! \cdot \! k) f^2  \,,\\ 
    H &= 3|f|^2 + (n^2)^2 k^2 - (n \! \cdot \! k) (f + f^*) n^2 \, ,
\end{align}
and ${}^*$ denotes complex conjugation. We have also dropped any subleading $\varepsilon$ terms that do not contribute as $\varepsilon \to 0$. 

We then proceed to evaluate the difference between the correlators
\begin{align}
g_E^{\rm Q}(t) &= g^2 \left\langle {\rm Tr}_c \left( U_{[-\infty,t]} F_{0i}(t) U_{[t,0]} F_{0i}(0) U_{[0,-\infty]} \right) \right\rangle = g^2 \left\langle {\rm Tr}_c \left( U_{[-\infty,t]} F_{0i}(t) U_{[t,+\infty]} U_{[+\infty,0]} F_{0i}(0) U_{[0,-\infty]} \right) \right\rangle \,, \\
g_E^{\rm Q\bar{Q}}(t) &= g^2 T_F \left\langle  \left( {F}_{0i}^a(t) \ml{W}_{[t, 0]}^{ab}
   {F}_{0i}^b(0) \right) \right\rangle = g^2 T_F \left\langle  \left( {F}_{0i}^a(t) \ml{W}_{[t, +\infty]}^{ac} \ml{W}_{[+\infty, 0]}^{cb}
   {F}_{0i}^b(0) \right) \right\rangle \,,
\end{align}
where the last arrangement of operators in each correlator is the one that is better suited for direct evaluation on the Schwinger-Keldysh contour, because in each case we have anti-time ordered operators grouped together on the left and time-ordered operators grouped on the right regardless of whether the time $t$ satisfies $t>0$ or $t<0$. Concretely, in both correlators, we take operators on the left of the Wilson lines extending from $t$ to $t=+\infty$ (including these Wilson lines) to be of type $2$, and operators on the right to be of type $1$.

\begin{figure}
    \begin{subfigure}[t]{0.49\textwidth}
        \centering
        \includegraphics[height=2.2in]{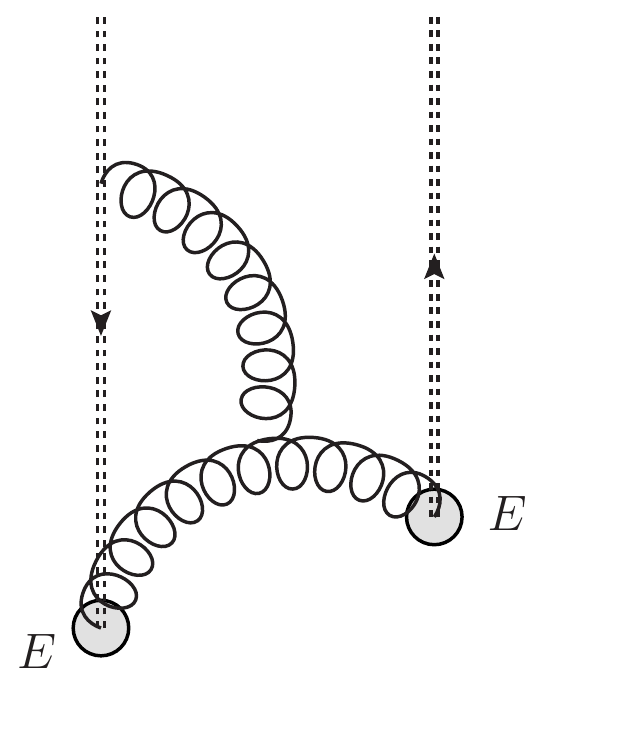}
        \caption{Quarkonium.}
        \label{fig:QQbar_EE}
    \end{subfigure}%
    ~
    \begin{subfigure}[t]{0.49\textwidth}
        \centering
        \includegraphics[height=2.2in]{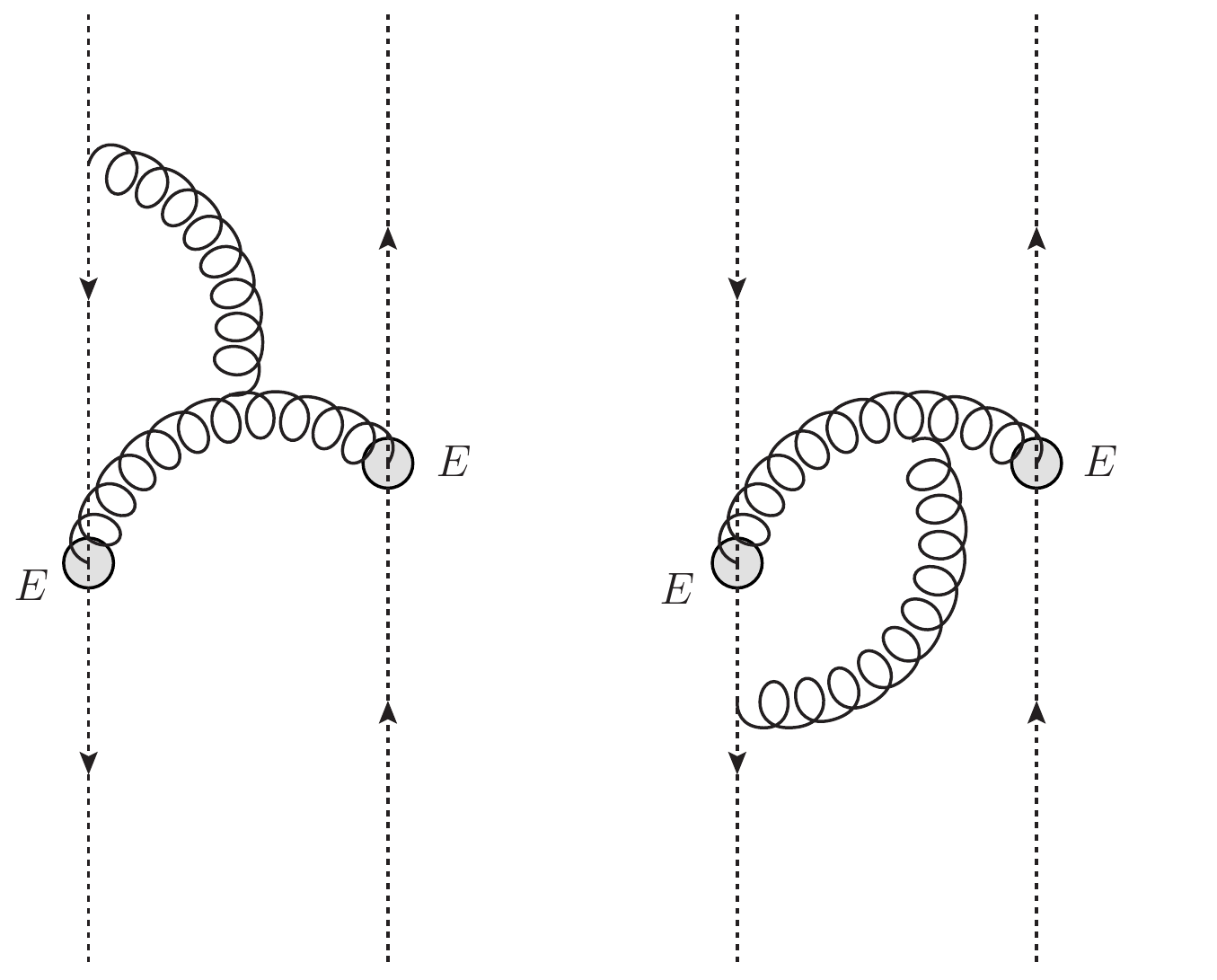}
        \caption{Heavy quark.}
        \label{fig:Q_EE}
    \end{subfigure}%
\caption{Feynman diagrams relevant for the difference between the chromoelectric field correlators for quarkonia (left) and heavy quarks (right). The blobs represent the chromoelectric fields while the double/single-dashed lines indicate the adjoint/fundamental Wilson lines. Similar diagrams where the gluon lines originating from the Wilson lines on the right are also included in the calculation.}
\label{fig:5}
\end{figure}

We perform a perturbative expansion on the coupling constant $g$ to calculate the difference
\be
g_E^{\rm Q\bar{Q}}(p_0) -  g_E^{\rm Q}(p_0) = \int_{-\infty}^{+\infty} \!\! \diff t \, e^{i p_0 t} \! \left(  g_E^{\rm Q\bar{Q}}(t) -  g_E^{\rm Q}(t) \right) \, .
\ee
In the language of Ref.~\cite{Binder:2021otw}, the difference comes solely from the diagrams of type $(5)$ and $(5r)$, i.e., diagrams with a triple-gauge boson vertex where only one of the three gluon lines is attached to a Wilson line. In the language of Refs.~\cite{Eller:2019spw,Burnier:2010rp}, the difference comes from the diagrams labeled as (j). These diagrams are shown in Fig.~\ref{fig:5}. Here we calculate the correlators directly as defined, rather than the time-ordered version as in the naive axial gauge calculation shown in the previous section. Following the calculation details of Ref.~\cite{Binder:2021otw}, we find these diagrams give
\begin{align}
    g_E^{\rm Q\bar{Q}}(p_0) -  g_E^{\rm Q}(p_0) = \int_{{\bs p}, k} & T_F g^4 N_c (N_c^2 - 1)  \pi \delta(k_0)  \big[ g_{\mu \nu} (p - 2k)_\rho + g_{\nu \rho} (k - 2p)_{\mu} + g_{\rho \mu} (p+k)_\nu \big] \nonumber \\ 
    \times &  (p_0 g_{i\rho'} - p_i g_{0\rho'} ) \big((p_0-k_0) g_{i\nu'} - (p_i - k_i) g_{0\nu} \big) \nonumber \\
    \times &  \Big( [D_{>}(p)]^{\rho' \rho} [D_T(p-k)]^{\nu \nu'} [D_T(k)]^{\mu 0} - [D_{\bar{T}}(p)]^{\rho' \rho} [D_>(p-k)]^{\nu \nu'} [D_>(k)]^{\mu 0} \nonumber \\
    & \,\,\,  - [D_{T}(p)]^{\rho \rho' } [D_>(p-k)]^{\nu' \nu } [D_>(k)]^{ 0 \mu}  + [D_{>}(p)]^{\rho \rho' } [D_{\bar{T}}(p-k)]^{\nu' \nu } [D_{\bar{T}}(k)]^{0 \mu} \Big) \, .
\end{align}
Here we have taken the color contractions out of the propagators and therefore only spacetime indices remain. The Dirac delta $\delta(k_0)$ appears when taking the difference between contributions with different pole prescriptions. These prescriptions are determined by the Wilson line regulator $\eta$ (see main text around~\eqref{eq:Wilson-line-regulated}) in each of the two correlators. In fact, this is the only difference between the expressions resulting from the Feynman rules for each diagram. It results in contributions of the form
\be
\frac{1}{k_0 + i\eta} - \frac12 \left( \frac{1}{k_0 + i \eta} + \frac{1}{k_0 - i \eta} \right) = \frac12 \left(\frac{1}{k_0 + i \eta} - \frac{1}{k_0 - i \eta} \right) \overset{\eta \to 0}{\longrightarrow} - i \pi \delta(k_0) \nonumber \, ,
\ee
where the first term can be traced back to an adjoint Wilson line attached to $t = +\infty$ (as in the quarkonium correlator), and the second term is the sum of the contributions from a fundamental Wilson line attached to $t=+\infty$ and another one attached to $t=-\infty$ (as in the heavy quark correlator).

The rest of the calculation is tedious, but straightforward. The most sensitive terms come from the propagator that connects the triple-gauge boson vertex with the Wilson line. We list them here explicitly:
\begin{align}
    \left. [D_T(k)]^{\mu 0} \right|_{k_0=0} &= \frac{-i}{{\bs k}^2-i\varepsilon} \left[ -g^{\mu 0} + \frac{i\lambda {\bs k}{}^2 k^{\mu} }{\lambda^2 {({\bs k}{}^2)}{}^2 + (1 - 2\lambda^2 {\bs k}{}^2 ) i \varepsilon } \right] \,, \\
    \left. [D_{\bar{T}}(k)]^{0 \mu} \right|_{k_0=0} &= \frac{i}{{\bs k}^2+i\varepsilon} \left[ -g^{ 0 \mu} + \frac{-i\lambda {\bs k}{}^2 k^{\mu} }{\lambda^2 {({\bs k}{}^2)}{}^2 - (1 - 2\lambda^2 {\bs k}{}^2 ) i \varepsilon } \right]  \,, \\
    \left. [D_>(k)]^{\mu 0} \right|_{k_0=0} &= \frac{\varepsilon}{{({\bs k}{}^2)}{}^2 + \varepsilon^2}  \left[ -g^{\mu 0} + \frac{i \big[ 2\lambda^3 {({\bs k}{}^2)}{}^3 - \lambda {({\bs k}{}^2)}{}^2 \big] k^{\mu} }{\lambda^4 {({\bs k}{}^2)}{}^4 + (1 - 2\lambda^2 {\bs k}{}^2 ){}^2 \varepsilon^2 } \right]  \,,\\
    \left. [D_>(k)]^{0 \mu} \right|_{k_0=0} &= \frac{\varepsilon}{{({\bs k}{}^2)}{}^2 + \varepsilon^2}  \left[ -g^{0 \mu} + \frac{-i \big[2\lambda^3 {({\bs k}{}^2)}{}^3 - \lambda {({\bs k}{}^2)}{}^2 \big] k^{\mu} }{\lambda^4 {({\bs k}{}^2)}{}^4 + (1 - 2\lambda^2 {\bs k}{}^2 ){}^2 \varepsilon^2 } \right] \, .
\end{align}
Here we have used that, as a distribution acting on continuous functions, $\delta(k_0) \Theta(k_0) = \frac12 \delta(k_0)$. Performing the index contractions leads to various integral structures. After isolating the contributions where the ${\bs k}$ momentum flowing in the propagators decouples from ${\bs p}$, using $\int \frac{\diff^d {\bs k} }{(2\pi)^d} {(({\bs k})^{2})}^{n} = 0$  for any integer $n$ in dimensional regularization for $d = 3 -\tilde{\epsilon}$, and using the symmetries of the integrand, one can reduce the expression for the difference to
\begin{align}
    g_E^{\rm Q\bar{Q}}(p_0) -  g_E^{\rm Q}(p_0) = \, & T_F g^4 N_c (N_c^2-1) \int_{{\bs p},k}  \frac{(2\pi)^2 \delta(k_0) \delta(p^2)}{((p_0^2 - ({\bs p} - {\bs k})^2 )^2 + \varepsilon^2) ( ({\bs k}{}^2)^2 + \varepsilon^2 )}  \big( p_0^2 - ({\bs p} - {\bs k})^2 \big)   \\
    & \times \bigg[   2 (1-d) {\bs k}^2 p_0^3  - \frac{ 2 (d-1) p_0^2 + {\bs k}^2 }{2} \bigg( \frac{\varepsilon \lambda^3 ({\bs k}{}^2)^4 }{\lambda^4 ({\bs k}{}^2)^4 + (1 - 2\lambda^2 {\bs k}{}^2 )^2 \varepsilon^2} - \frac{\varepsilon^3 \lambda {\bs k}{}^2 (1-2\lambda^2 {\bs k}{}^2 ) }{\lambda^4 ({\bs k}{}^2)^4 + (1 - 2\lambda^2 {\bs k}{}^2 )^2 \varepsilon^2} \bigg)  \bigg]\,, \nonumber
\end{align}
where we have assumed $p_0>0$ and taken $\frac{\varepsilon}{{(p^2)}^2 + \varepsilon^2} \to \pi \delta(p^2)$ from the start since this does not cause any singularity in the calculation. 
The first term (i.e., the ${\bs k}^2$ term) in the square bracket gives the Feynman gauge result. The other two terms, together with the overall factor $( ({\bs k}{}^2)^2 + \varepsilon^2 )^{-1}$, give a contribution that is smaller than
$ {\bs k}{}^2 \delta({\bs k}^2)$ in the limit $\varepsilon\to0$, which is automatically vanishing. This can be seen by noting that
\begin{align}
    & \frac{1}{({\bs k}{}^2)^2 + \varepsilon^2} \frac{\varepsilon \lambda^2 ({\bs k}{}^2)^4 }{\lambda^4 ({\bs k}{}^2)^4 + (1 - 2\lambda^2 {\bs k}{}^2 )^2 \varepsilon^2} \leq \frac{\varepsilon \lambda^2 ({\bs k}{}^2)^2 }{\lambda^4 ({\bs k}{}^2)^4 + (1 - 2\lambda^2 {\bs k}{}^2 )^2 \varepsilon^2} \overset{\varepsilon \to 0}{\longrightarrow} \pi \lambda^2 ({\bs k}{}^2)^2 \delta( \lambda^2 ({\bs k}{}^2)^2 ) = \frac{\pi {\bs k}{}^2 }{2}  \delta({\bs k}{}^2) \nonumber \, , \\
    & \frac{1}{({\bs k}{}^2)^2 + \varepsilon^2} \frac{\varepsilon^3  {\bs k}{}^2 |1-2\lambda^2 {\bs k}{}^2 | }{\lambda^4 ({\bs k}{}^2)^4 + (1 - 2\lambda^2 {\bs k}{}^2 )^2 \varepsilon^2} \leq \frac{\varepsilon}{({\bs k}{}^2)^2 + \varepsilon^2} \frac{ {\bs k}{}^2  }{|1 - 2\lambda^2 {\bs k}{}^2|} \overset{\varepsilon \to 0}{\longrightarrow} \frac{\pi {\bs k}{}^2  }{|1 - 2\lambda^2 {\bs k}{}^2|} \delta({\bs k}{}^2) = \pi {\bs k}{}^2 \delta({\bs k}{}^2) \nonumber \, ,
\end{align}
in tandem with the fact that the rest of the integral is non-singular at ${\bs k} = 0$ because the possible divergence in the principal value $\mathcal{P} \big[ (p_0^2 - ({\bs p} - {\bs k})^2 )^{-1} \big] $ is tamed by the ${\bs k}{}^2$ factor in the integral measure. Thus, at any finite $\lambda$, only the term that corresponds to the Feynman gauge calculation remains, and the result is independent of $\lambda$. Therefore, by taking $\eta\to0$ and evaluating the integrals first, we find that the limit $\lambda\to0$ gives the same result as the other gauges.

For completeness, we evaluate the remaining (finite) piece that gives the gauge invariant result for the difference. Now we can set $d=3$ because the integrals are strictly convergent and no regularization is required. The result is
\begin{align}
    g_E^{\rm Q\bar{Q}}(p_0) -  g_E^{\rm Q}(p_0) &= T_F g^4 N_c (N_c^2-1) \int_{{\bs p},k}  (2\pi)^2 \delta(k_0) \delta(p^2)  \mathcal{P} \left( \frac{ - 2 (d-1) p_0^3 }{(p_0^2 - ({\bs p} - {\bs k})^2 ) {\bs k}{}^2} \right) \nonumber \\
    &= \frac{T_F g^4 N_c (N_c^2-1)}{(2\pi)^3} \left( - 8 p_0^3 \right) \int_0^\infty \diff|{\bs p}| |{\bs p}|^2 \delta(p_0^2 - |{\bs p}|^2) \int_0^{\infty} \diff|{\bs k}| |{\bs k}|^2 \int_{-1}^1 \frac{\diff u}{ {\bs k}{}^2 (2 |{\bs p}| |{\bs k}| u - {\bs k}{}^2 ) } \nonumber \\
    &= \frac{T_F g^4 N_c (N_c^2-1) p_0^3}{(2\pi)^3} (-2) \int_0^\infty \frac{\diff |{\bs k}|}{|{\bs k}|} \ln \left| \frac{1 - |{\bs k}|/(2p_0)}{1+|{\bs k}|/(2p_0) } \right| \nonumber \\
    &= \frac{T_F g^4 N_c (N_c^2-1) p_0^3}{(2\pi)^3} 2 \int_0^\infty \frac{\diff x}{x} \ln \left| \frac{1 + x}{1 - x } \right| \\\nn
    &= \frac{T_F g^4 N_c (N_c^2-1) p_0^3}{(2\pi)^3} \pi^2 \, ,
\end{align}
where we used our assumption $p_0>0$. Thus, we reproduced the difference observed in Ref.~\cite{Binder:2021otw} by using mixed axial-Feynman gauge.



\end{widetext}

\end{document}